\documentclass[conference,letterpaper]{IEEEtran}

\usepackage[utf8]{inputenc} 
\usepackage[T1]{fontenc}
\usepackage{url}
\usepackage{ifthen}
\usepackage{cite}
\usepackage[cmex10]{amsmath} 

\usepackage{amssymb}
\usepackage{graphicx}
\usepackage{caption}
\usepackage{subcaption}

\graphicspath{{figure/}}

\newcommand{\cb}[1]{$\text{CB}_{#1}$}
\newcommand{\ut}[1]{\mathbf{u}_{#1}}

\begin{document}
	\title{Partially Information Coupled Duo-Binary Turbo Codes}
	\author{\IEEEauthorblockN{Xiaowei Wu, Min Qiu, and Jinhong Yuan}
		\IEEEauthorblockA{School of Electrical Engineering and Telecommunications\\The University of New South Wales, Sydney, Australia\\Email: xiaowei.wu@unsw.edu.au, min.qiu@unsw.edu.au, j.yuan@unsw.edu.au}
	}	
	\maketitle
	
	\begin{abstract}
		Partially information coupled turbo codes (PIC-TCs) is a class of spatially coupled turbo codes that can approach the BEC capacity while keeping the encoding and decoding architectures of the underlying component codes unchanged. However, PIC-TCs have significant rate loss compared to its component rate-$\frac{1}{3}$ turbo code, and the rate loss increases with the coupling ratio. 
		To absorb the rate loss, in this paper, we propose the partially information coupled duo-binary turbo codes (PIC-dTCs). 
		Given a rate-$\frac{1}{3}$ turbo code as the benchmark, we construct a duo-binary turbo code by introducing one extra input to the benchmark code. Then, parts of the information sequence from the original input are coupled to the extra input of the succeeding code blocks. 
		By looking into the graph model of PIC-dTC ensembles, we derive the exact density evolution equations of the PIC-dTC ensembles, and compute their belief propagation decoding thresholds on the binary erasure channel. 
		Simulation results verify the correctness of our theoretical analysis, and also show significant error performance improvement over the uncoupled rate-$\frac{1}{3}$ turbo codes and existing designs of spatially coupled turbo codes. 
	\end{abstract}
	
	\section{Introduction}
	Spatial coupling is a class of code construction techniques which connects a sequence of component codes to form a long codeword chain. It was firstly introduced in \cite{Felstrom-convLDPC} for constructing convolutional low-density parity-check (LDPC) codes, which is also known as spatially coupled LDPC (SC-LDPC) codes. 
	Extensive research has been conducted on SC-LDPC codes (see \cite{Kudekar-scLDPC,Lentmaier-scLDPC,Xie-scLDPC} and the references therein) since then. 
	In addition, the spatial coupling technique has also been applied to turbo-like codes (SC-TCs), including parallel and serially concatenated convolutional codes \cite{Moloudi-scTurbo}, hybrid concatenated convolutional codes \cite{Moloudi-scHCC}, laminated turbo codes \cite{Huebner-LaminatedTC}, and braided convolutional codes \cite{Zhang-bcc}. 
	All the above works have reported that the spatially coupled codes can provide close-to-capacity performance and perform much better than their uncoupled counterparts. 	 
	Most notably, it has been theoretically proven in \cite{Kudekar-scLDPC} that the belief propagation (BP) decoding thresholds of SC-LDPC codes can converge to the maximum-a-posteriori (MAP) decoding thresholds of their underlying component codes, namely the threshold saturation phenomenon. 
	 In \cite{Moloudi-scTurbo}, the authors also proved that their proposed SC-TCs have threshold saturation. Apart from the performance improvement, another advantage of spatially coupled codes is that they can be decoded by a windowed decoder with lower decoding latency compared to decoding the whole codeword block. 
	 Due to these reasons, it is believed that spatially coupled codes would have a wide range of applications in communications.
	
	In this work, we focus on a particular spatial coupling technique, namely the partially information coupling (PIC). PIC was firstly introduced in \cite{Yang-picTurbo} with LTE turbo codes as component codes (PIC-TCs).
	In a PIC code, consecutive component code blocks (CBs) are coupled by sharing a portion of the information bits between each other. 
	Apart from turbo codes, PIC can be directly applied to various types of component codes, such as LDPC codes \cite{Yang-PIC_LDPC} and polar codes \cite{Wu-picPolar,Wu-picPolarBICM}, without changing the encoding and decoding architectures of the underlying component codes. 
	However, due to the coupling, (i.e., the same information sharing between CBs), PIC codes have significant rate loss compared to its component codes. For example, in \cite{Yang-picTurbo,Qiu-picTurboDE}, with a rate-$\frac{1}{3}$ turbo code as component code, the PIC-TC ensembles have a code rate of $\frac{1}{5}$ when half of the information bits are coupled.
	
	In this paper, we propose a new class of PIC codes, namely the partially information coupled duo-binary turbo codes (PIC-dTCs). Such codes do not have the rate-loss as appeared in the conventional PIC-TCs. More importantly, they can retain the same rate regardless of the couple-ratio.	
	Given a $\textnormal{rate-}\frac{1}{3}$ benchmark turbo code, i.e., a parallel concatenation of two rate-$\frac{1}{2}$ recursive systematic convolutional (RSC) code, we construct a duo-binary turbo code \cite{Berrou-nonBinaryTurbo,Berrou-duoBinaryTurbo} by introducing one extra input to the component RSC code. 
	After that, we apply partially information coupling to a shortened duo-binary turbo code. The resultant PIC-dTCs have the same code rate as the benchmark turbo code. 	
	We study the performance of the PIC-dTCs over the binary erasure channel (BEC). We first look into the graph model of the code ensembles. Based on the graph representation, we derive the exact density evolution (DE) equations for the PIC-dTC ensembles with any given coupling ratio and coupling memory. The DE analysis shows that our codes are within a gap of 0.001 to the BEC capacity . 
	Simulation results confirm our theoretical analysis, and also show significant error performance improvement over the benchmark turbo code and existing designs of spatially coupled turbo codes.
	
	\section{Partially Information Coupled Duo-Binary Turbo Codes}
	In this section, we introduce the architecture of PIC-dTCs. We first present the construction of the uncoupled duo-binary turbo codes. Then, we describe the encoding of PIC-dTCs with coupling memory $m=1$, i.e., a CB only shares information with between two consecutive CBs. After that, we will give a general description on the encoding and decoding of PIC-dTCs with coupling memory $m \geq 1$. 
	
	\subsection{Construction of Duo-Binary Turbo Codes}
	We consider a rate-$\frac{1}{3}$ turbo code (referred to as TC1), which is a parallel concatenation of two identical rate-$\frac{1}{2}$ RSC code (referred to as RSC1). To build PIC-dTCs from TC1, the first step is to construct a rate-$\frac{1}{2}$ duo-binary turbo code (referred to as TC2), which is a parallel concatenation of two identical rate-$\frac{2}{3}$ RSC code (referred to as RSC2).
	
	Let $\mathbf{g}_f$ and $\mathbf{g}_b$ denote the forward and feedback generator polynomial of RSC1, respectively. The generator matrix of RSC1 can then be written as 
	\begin{align}
	\mathbf{G}_{\text{RSC}1}=\left [{\begin{matrix} 1  ~ \frac{\mathbf{g}_f}{\mathbf{g}_b}  \end{matrix}}\right ].
	\end{align}
	Let $\ut{}$ denote an information sequence of length $K$. 
	When we encode $\ut{}$ with RSC1, the output is ${\ut{} \cdot \mathbf{G}_{\text{RSC}1}=\left [{\begin{matrix} \ut{} ~ \mathbf{v} \end{matrix}}\right ]}$, where ${\mathbf{v} = \frac{\ut{} \cdot \mathbf{g}_f}{\mathbf{g}_b}}$ is the parity sequence.
	
	By introducing one extra input $\ut{}'$ to RSC1, the generator matrix of RSC2 is in the form of
	\begin{align}
	\mathbf{G}_{\text{RSC}2}=\left [{\begin{matrix} 1 & 0 & \frac{\mathbf{g}_f}{\mathbf{g}_b}  \\ 0 & 1 & \frac{\mathbf{g}_f'}{\mathbf{g}_b}  \end{matrix}}\right ],
	\end{align}
	where $\mathbf{g}_f$ and $\mathbf{g}_f'$ are the forward generator polynomial of the two information sequences $\ut{}$ and $\ut{}'$, respectively, and ${\mathbf{g}_f \neq \mathbf{g}_f'}$. 
	Both information sequences share the same feedback generator polynomials $\mathbf{g}_b$ so that RSC1 and RSC2 have the same number of states. 
	The encoder output of RSC2 is computed as ${\ut{} \cdot \mathbf{G}_{\text{RSC}2}=\left [{\begin{matrix} \ut{} ~ \ut{}' ~ {\mathbf{v}} \end{matrix}}\right ]}$, where ${\mathbf{v}=\frac{\ut{} \cdot \mathbf{g}_f}{\mathbf{g}_b}+\frac{\ut{}' \cdot \mathbf{g}_f'}{\mathbf{g}_b}}$.
	This ensures RSC1 can be obtained by shortening RSC2, i.e., when $\ut{}'=\mathbf{0}$, the parity sequence from RSC2 is the same as that of RSC1. Consequently, TC1 can be obtained by shortening TC2 in the same manner. 
	
\begin{figure}[t]
		\centering
		\includegraphics[width=0.35\textwidth]{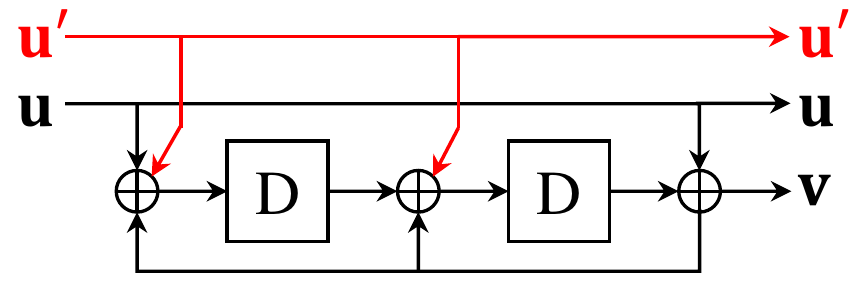}	
		\caption{Block diagram of the rate-$\frac{2}{3}$ RSC encoder. }
		\label{fig:RSC_enc}	
	\end{figure}

	An example is shown in Fig. \ref{fig:RSC_enc}, where ${\mathbf{G}_{\text{RSC}1}=\left [{\begin{matrix} 1  ~ \frac{5}{7} \end{matrix}}\right ]}$, and ${\mathbf{G}_{\text{RSC}2}=\left [{\begin{smallmatrix} 1 & 0 & \frac{5}{7} \\ 0 & 1 & \frac{3}{7} \end{smallmatrix}}\right ]}$, respectively. Here, ${\mathbf{g}_f'=3}$ is obtained by exhaustive search to ensure that RSC2 has a good distance spectrum, and the resultant TC2 has a good decoding threshold. The extra input $\ut{}'$ are highlighted in red. 
	
	\subsection{Encoding of PIC-dTCs with Coupling Memory $m=1$}	
\begin{figure}[t]
		\centering
		\includegraphics[width=0.4\textwidth]{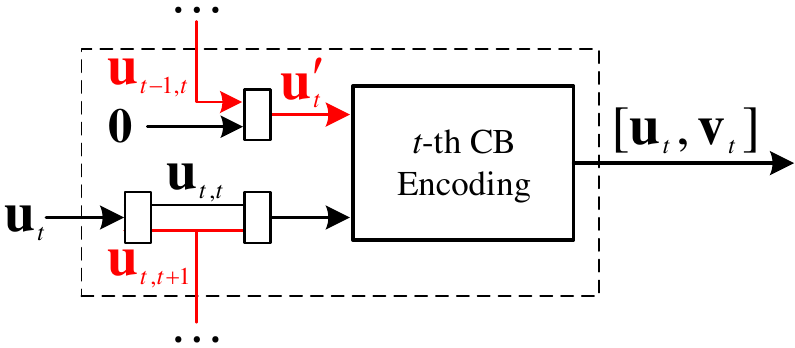}	
		\caption{Block diagram of PIC-dTC encoder at time $t$. }
		\label{fig:TB_encode}	
	\end{figure}

	In the PIC-dTC encoding process, the information sequence $\mathbf{u}$ is divided into $L$ sub-sequences ${\ut{1}, \hdots, \ut{L}}$. For the time instance ${t=1, \hdots, L}$, let \cb{t} represent the $t$-th CB. 
	The block diagram of the PIC-dTC encoder at time $t$ with coupling memory ${m=1}$ is depicted in Fig. \ref{fig:TB_encode}. 
	The CB encoder takes the $t$-th sub-sequence, i.e., $\ut{t}$, as the first input sequence, and takes ${\ut{t}'=[\ut{t-1,t}, \mathbf{0}]}$ as the second input sequence. Here, $\ut{t-1,t}$ is the coupled information sequence shared between \cb{t} and \cb{t-1}, and shortened bits $\mathbf{0}$ are inserted in the second input sequence so that the length of $\ut{t}$ and $\ut{t}'$ are equal. 
	Then, $\ut{t}$ is decomposed into $\ut{t,t}$ and $ \ut{t,t+1}$, where $\ut{t,t}$ represents the uncoupled information sequence, i,e. the information only stays in \cb{t}, and $\ut{t,t+1}$ represents the coupled information sequence  shared between \cb{t} and \cb{t+1}. 
	After CB encoding, the codeword of \cb{t} is obtained as ${[\ut{t},\mathbf{v}_t]}$, where $\mathbf{v}_t$ is the parity sequence. 
	Note that $\ut{t-1,t}$ is not included in the codeword of \cb{t} because it is included in the codeword of \cb{t-1}. 
	To terminate the coupling, both ends of the coupling chain are set to zero, i.e., at time ${t=1}$, ${\ut{0,1}=\mathbf{0}}$, and at time ${t=L}$, ${\ut{L,L+1}=\mathbf{0}}$. 
	
	In \cb{t}, the information length is ${K=\|\ut{t} \|=\|\ut{t}' \|}$, the  length of coupled sequence is ${K_c=\|\ut{t-1,t} \|=\|\ut{t,t+1} \|}$, and the length of parity sequence is ${2K=\|\mathbf{v}_t\|}$.  
	The code rate of PIC-dTC is 
	\begin{align}
	R  = \frac{KL-K_c}{3KL-K_c} = \frac{L-\lambda}{3L-\lambda} \overset{L\rightarrow\infty}{=} \frac{1}{3}.
	\end{align}
	We define ${\lambda\triangleq\frac{K_c}{K}}$ as the coupling ratio, and $\lambda\in [0,1]$. It is an important parameter that determines the decoding threshold, which will be discussed later. 
		
	
	\subsection{Encoding of PIC-dTCs with Coupling Memory $m \geq 1$}
	For coupling memory ${m \geq 1}$, \cb{t} are coupled with $m$ preceding CBs (from \cb{t-m} to \cb{t-1}) and $m$ succeeding CBs (from \cb{t+1} to \cb{t+m}). The length of coupled sequence becomes $\frac{K_c}{m}$.
	Specifically, at time $t$, the CB encoder takes $\ut{t}$ as the first input sequence, and ${\ut{t}'=[\ut{t-1,t}, \hdots, \ut{t-m,t}, \mathbf{0}]}$ as the second input sequence. 
	In the mean time, $\ut{t}$ is decomposed into ${\ut{t,t}, \ut{t,t+1}, \hdots, \ut{t,t+m}}$, where ${\ut{t,t+1}, \hdots, \ut{t,t+m}}$ are passed to $m$ succeeding CBs, respectively. 
	 After CB encoding, the codeword of \cb{t} is obtained as
	${[\ut{t},\mathbf{v}_t]}$, i.e.,	
	${\ut{t-1,t}, \hdots, \ut{t-m,t}}$ is not included in the codeword. 
	The code rate of PIC-dTC is 
	\begin{align}
	R  = \frac{KL-\sum_{i=1}^{m} \frac{i \cdot K_c}{m}}{3KL-
		\sum_{i=1}^{m}  \frac{i \cdot K_c}{m}} 
	= \frac{L-  \frac{m+1}{2} \lambda } {3L-  \frac{m+1}{2} \lambda} 
	\overset{L\rightarrow\infty}{=} \frac{1}{3}.
	\end{align}
	
	\textbf{Remark}: We compare the construction of the PIC-dTCs with that of the PIC-TCs \cite{Yang-picTurbo,Qiu-picTurboDE}. For the PIC-TCs with TC1 as component code, the CB encoder only has one information sequence. Hence, there is a rate loss related to the coupling ratio $\lambda$. Specifically, the code rate of PIC-TCs is given by ${\frac{1-\lambda }{3-\lambda}}$, where ${\lambda\in[0,\frac{1}{2}]}$. 
	Puncturing is required in order to increase the code rate of PIC-TCs to $\frac{1}{3}$. 
	
	For the PIC-dTCs, the CB encoder takes two input sequences $\ut{t}$ and $\ut{t}'$. Unlike the PIC-TCs, we add an extra input to the TC1, resulted in TC2, and put the coupled sequence from previous CBs into the extra input sequence. This design can absorb the rate loss as appeared in the PIC-TCs. Hence, it allows the resultant PIC-dTCs to maintain the rate of TC1, i.e., 1/3, without rate loss. In addition, the PIC-dTCs can attain the close-to-capacity performance which will be shown in our performance analysis and numerical results later.
	
	\subsection{Decoding of PIC-dTCs}
	The decoding of PIC-dTCs can be accomplished by a feed-forward and feed-back (FF-FB) scheme \cite{Yang-picTurbo,Qiu-picTurboDE} in an iterative manner. In short, it employs a serial scheduling by decoding from the first CB to the last CB serially and then decoding backwards from the last CB to the first CB if necessary. For \cb{t}, the decoder takes the received signal associated with the codeword ${[\ut{t},\mathbf{v}_t]}$ as well as the extrinsic information associated with ${\ut{t-m,t}, \hdots, \ut{t-1,t}}$ and ${\ut{t,t+1}, \hdots, \ut{t,t+m}}$  as input, and outputs the soft-decision estimation of the information sequences. The decoding of RSC2 is realized by the BCJR algorithm \cite{bcjr}\cite{turbo_yuan}. 
	
	\section{Performance Analysis of PIC-{\upshape d}TCs}
	In this section, we analyse the decoding performance of the PIC-dTCs by using density evolution. We first look into the corresponding graph model of the PIC-dTC. Later, the exact DE equations are then derived based on the graph model.
	
	\subsection{Graph Model Representation}
\begin{figure}[t]
		\centering
		\hspace{-2mm}
		\subcaptionbox{\label{fig:factor_graph_uc}}
		{\includegraphics[height=4.1cm]{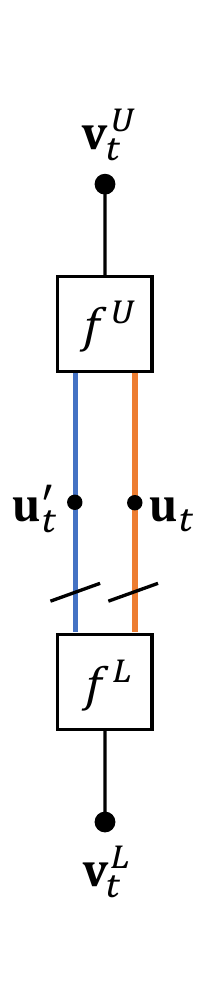}}
		\hspace{-4mm}
		\subcaptionbox{\label{fig:factor_graph_1}}
		{\includegraphics[height=4.1cm]{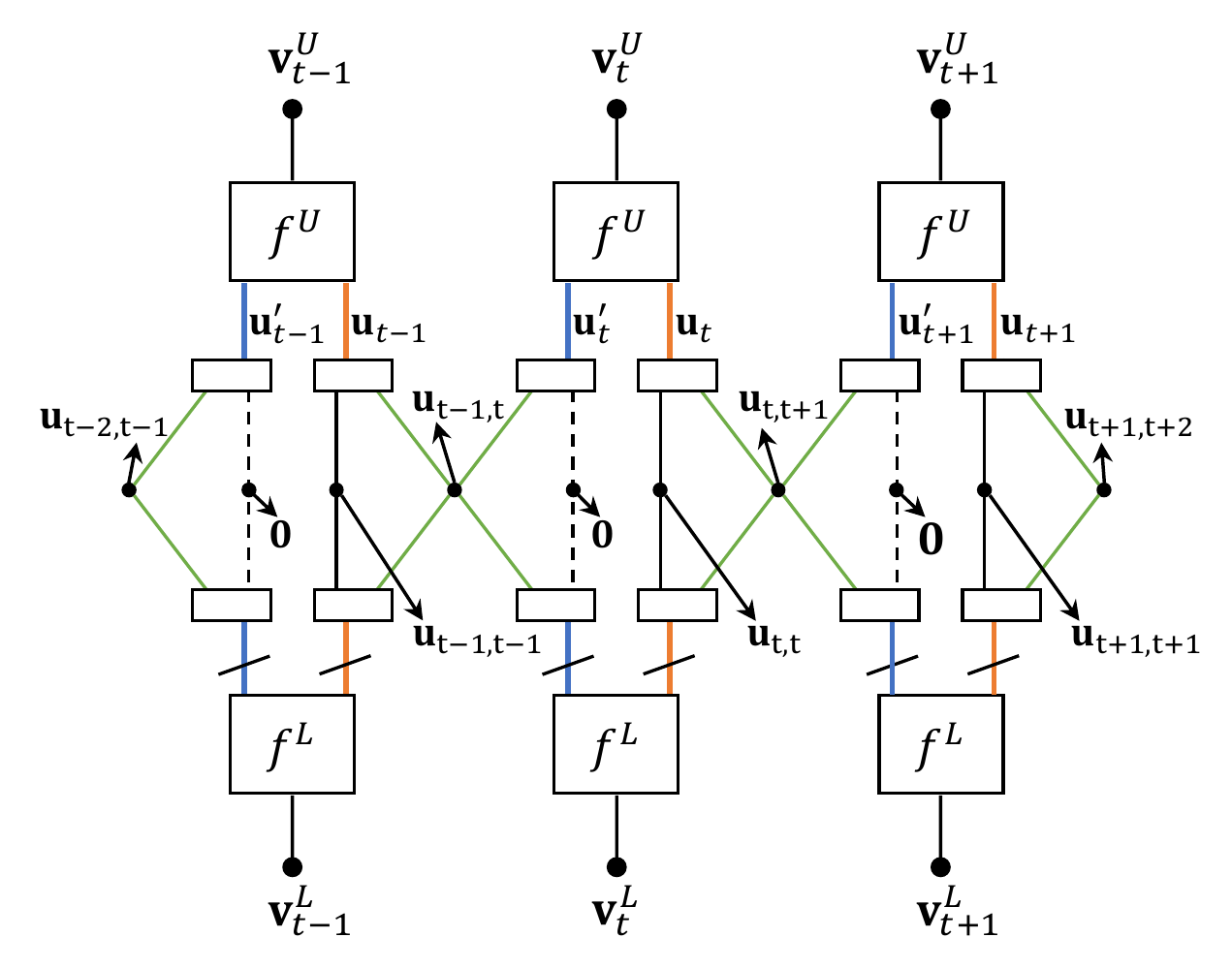} } 
		\hspace{-5mm}
		\subcaptionbox{ \label{fig:factor_graph_m}}
		{\includegraphics[height=4.1cm]{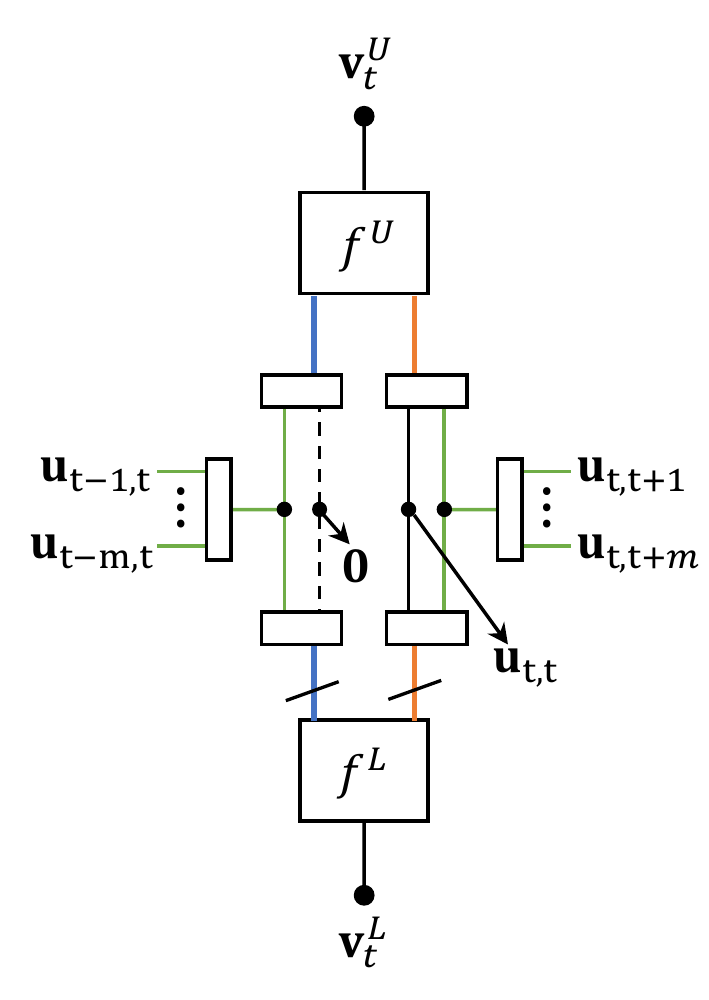}	}
		\caption{Compact factor graph of (a) uncoupled turbo codes, and (b) PIC-dTC with coupling memory ${m=1}$, and (c) PIC-dTC with coupling memory ${m \geq 1}$.}
		\label{fig:factor_graph}	
	\end{figure}

	For the ease of presentation, we first consider the coupling memory ${m = 1}$ and then extend to the coupling memory ${m \geq 1}$.
	Fig. \ref{fig:factor_graph_uc} shows the compact graph representation \cite{Moloudi-scTurbo} of the uncoupled TC2. In the compact graph, at time $t$, the two information sequences $\ut{t}$ and $\ut{t}'$ are represented by two different variable nodes. We use two factor nodes $f^U$ and $f^L$ to represent the upper and the lower RSC decoders, respectively. 
	Each of the variable nodes is connected with the upper and lower factor nodes, meaning that the extrinsic information is passed between the upper and lower decoder via the corresponding variable node. 
	The parity sequences enter the upper and lower decoder are denoted by $\mathbf{v}^U_t$ and  $\mathbf{v}^L_t$, respectively. The interleaver is represented by a slash on the edge between the variable nodes and the lower factor node. 
	
	Fig. \ref{fig:factor_graph_1} shows the compact graph of PIC-dTC with coupling memory ${m=1}$. For each CB, we treat the uncoupled information, the coupled information, and the shortened bits node (denoted by $\mathbf{0}$ in	Fig. \ref{fig:factor_graph_1}) separately. At time $t$, we use two variable nodes to represent $\ut{t}'$, i.e., we treat $\ut{t-1,t}$ and $\mathbf{0}$ as two separate nodes. Also, we use two variable nodes to represent $\ut{t}$ by treating $\ut{t,t}$ and $\ut{t,t+1}$ separately. 
	As $\ut{t-1,t}$ is shared by $\ut{t-1}$ and $\ut{t}'$, the node representing $\ut{t-1,t}$ is connected to both \cb{t-1} and \cb{t}, meaning that $\ut{t-1,t}$ is encoded twice at time ${t-1}$ and time $t$. 
	Due to the similar reason, the node $\ut{t,t+1}$ is connected to both \cb{t} and \cb{t+1}. 
	
	In summary, the compact graph shows that the extrinsic information of $\ut{t-1,t}$ is passed between the upper and lower factor nodes of \cb{t-1} and \cb{t}; the extrinsic information of $\ut{t,t+1}$ is passed between the upper and lower factor nodes of \cb{t} and \cb{t+1}; and the extrinsic information of $\ut{t,t}$ is passed between the upper and lower factor nodes of \cb{t} only. 

	\subsection{Density Evolution for Coupling Memory $m=1$} \label{sec:de}
	For transmission over the BEC, the asymptotic behaviour of the PIC-dTCs can be analysed by tracking the evolution of the erasure probability over decoding iterations. Here, we first consider coupling memory ${m=1}$. 
	
	Let $\varepsilon$ denote the channel erasure probability.
	At time $t$ and the $i$-th iteration, let $\bar{p}_{1,L,t}^{(i)} $ and $\bar{p}_{2,L,t}^{(i)}$ denote the average extrinsic erasure probability passed from $\ut{t}$ and $\ut{t}'$ to $f^U$, respectively. Let $p_{1,U,t}^{(i)} $ and $p_{2,U,t}^{(i)}$ denote the extrinsic erasure probability from $f^U$ to $\ut{t}$ and $\ut{t}'$, respectively.	
	Likewise, let $\bar{p}_{1,U,t}^{(i)} $ and $\bar{p}_{2,U,t}^{(i)}$ denote the average extrinsic erasure probability passed from $\ut{t}$ and $\ut{t}'$ to $f^L$, respectively. Let $p_{1,L,t}^{(i)} $ and $p_{2,L,t}^{(i)}$ denote the extrinsic erasure probability from $f^L$ to $\ut{t}$ and $\ut{t}'$, respectively.	
	
	Based on the graph model in Fig. \ref{fig:factor_graph_1}, $\bar{p}_{1,L,t}^{(i)}$ is the weighted sum of the erasure probabilities of variable node $\ut{t,t}$ and $\ut{t,t+1}$, and $\bar{p}_{2,L,t}^{(i)}$ is the weighted sum of the erasure probabilities of variable node $\ut{t-1,t}$ and $\mathbf{0}$. 
	For $\bar{p}_{1,L,t}^{(i)}$, $\ut{t,t}$ can only obtain extrinsic information from $f^L$ at time $t$, while $\ut{t,t+1}$ can obtain extrinsic information from $f^L$ at time $t$, as well as from $f^L$ and $f^U$ at time $t+1$. It is computed as	
	\vspace{-0.5mm}
	\begin{align}
	\bar{p}_{1,L,t}^{(i)}=\varepsilon \cdot p_{1,L,t}^{(i)} \cdot \left( 1 - \lambda + \lambda \cdot p_{2,U,t+1}^{(i-1)} \cdot  p_{2,L,t+1}^{(i)}   \right) .
	\vspace{-0.5mm}
	\end{align}
	For $\bar{p}_{2,L,t}^{(i)}$, the erasure rate of $\mathbf{0}$ is 0, while $\ut{t-1,t}$ can collect extrinsic information from $f^L$ at time $t$, as well as $f^U$ and $f^L$ at time ${t-1}$. It is computed as
	\vspace{-0.5mm}
	\begin{align}
	\bar{p}_{2,L,t}^{(i)}=\varepsilon \cdot \lambda \cdot  p_{2,L,t}^{(i)} \cdot p_{1,U,t-1}^{(i-1)}  \cdot  p_{1,L,t-1}^{(i)} .
	\end{align} \vspace{-5mm}
	$\!\!$Likewise, the average erasure probability from $\ut{t}$ and $\ut{t}'$ to $f^L$ is computed as
	\begin{align}
	\bar{p}_{1,U,t}^{(i)} = \varepsilon \cdot p_{1,U,t}^{(i)} \cdot  \left( 1 - \lambda + \lambda  \cdot p_{2,U,t+1}^{(i)} \cdot  p_{2,L,t+1}^{(i-1)} \right) , 
	\end{align}
	\vspace{-0.5mm}
	$\!\!$and 
	\vspace{-0.5mm}
	\begin{align}	
	\bar{p}_{2,U,t}^{(i)}=\varepsilon \cdot \lambda \cdot  p_{2,U,t}^{(i)}  \cdot  p_{1,U,t-1}^{(i)} \cdot p_{1,L,t-1}^{(i-1)} ,
	\end{align}
	respectively. 
	The erasure probability of $\mathbf{v}_t^U$ and $\mathbf{v}_t^L$ are the same as the channel erasure probability because no extrinsic information is passed to the parity nodes.
	
	Let $F_{1}^U$ and$F_{2}^U$ represent the information sequence transfer functions of $f^U$, and $F_{1}^L$ and $F_{2}^L$ represent the information sequence transfer functions of $f^L$, which are derived following \cite{Kurkoski-turboDE}, respectively. At time $t$ and the $i$-th iteration, the evolution of the erasure probability for $\ut{t}$ and $\ut{t}'$ inside $f^U$ is 
	\begin{align}
		p_{1,U,t}^{(i)}=F_{1}^U\left(\bar{p}_{1,L,t}^{(i)},\bar{p}_{2,L,t}^{(i)},\varepsilon\right),\\
		p_{2,U,t}^{(i)}=F_{2}^U\left(\bar{p}_{1,L,t}^{(i)},\bar{p}_{2,L,t}^{(i)},\varepsilon\right),
	\end{align}
	and the evolution of erasure probability inside $f^L$ is
	\begin{align}
	p_{1,L,t}^{(i)}=F_{1}^L\left(\bar{p}_{1,U,t}^{(i)},\bar{p}_{2,U,t}^{(i)},\varepsilon\right),\\
	p_{2,L,t}^{(i)}=F_{2}^L\left(\bar{p}_{1,U,t}^{(i)},\bar{p}_{2,U,t}^{(i)},\varepsilon\right).
	\end{align}
	The a-posteriori erasure probability of $\ut{t}$ after $i$ iterations is 
	\begin{align}
	p_{\ut{t}}^{(i)}=\varepsilon \!\cdot\! p_{1,U,t}^{(i)}  \!\cdot\! p_{1,L,t}^{(i)} \!\cdot\!\left(1-\lambda + \lambda \!\cdot\!  p_{2,U,t+1}^{(i)}  \!\cdot\! p_{2,L,t+1}^{(i)} \right). 
	\end{align}
	
	\subsection{Density Evolution for Coupling Memory $m \geq 1$}
	Here, we generalize the DE analysis to coupling memory ${m \geq 1}$.
	As shown in Fig. \ref{fig:factor_graph_m}, the erasure probability of $\ut{t}$ depends on the erasure probability of uncoupled information $\ut{t,t}$ as well as the coupled information ${\ut{t,t+1}, \hdots, \ut{t,t+m}}$. The average extrinsic erasure probability from $\ut{t}$ to the $f^U$ is computed as 
	\vspace{-2mm}
	\begin{align}
		\bar{p}_{1,L,t}^{(i)}=\varepsilon  \!\cdot\!  p_{1,L,t}^{(i)}  \!\cdot\!  \Big(  1 - \lambda + \frac{\lambda}{m} \sum_{j=1}^{m}  p_{2,U,t+j}^{(i-1)}   \!\cdot\!  p_{2,L,t+j}^{(i)}  \Big).
	\end{align}  
	
	The erasure probability of $\ut{t}'$ only depends on the erasure probability of the coupled information ${\ut{t-m,t}, \hdots, \ut{t-1,t}}$. The average extrinsic erasure probability from $\ut{t}'$ to the $f^U$ is computed as	
	\vspace{-2mm} 
	\begin{align}
	\bar{p}_{2,L,t}^{(i)}=\varepsilon \cdot  p_{2,L,t}^{(i)} \cdot \frac{\lambda}{m} \sum_{j=1}^{m} 
	 p_{1,U,t-j}^{(i-1)}  \cdot p_{1,L,t-j}^{(i)} .
	\end{align} 
	
	The average erasure probability from $\ut{t}$ and $\ut{t}'$ to $f^L$, i.e., $\bar{p}_{1,U,t}^{(i)}$ and $\bar{p}_{2,U,t}^{(i)}$, can be computed in similar manner, so we omit the details here. The a-posteriori erasure probability of $\ut{t}$ after $i$ iterations is 
	\vspace{-2mm}
	\begin{align}
	 \!  \!  \!  \!  p_{\ut{t}}^{(i)}=\varepsilon \!  \cdot  \! p_{1,U,t}^{(i)} \!  \cdot \!  p_{1,L,t}^{(i)}  \!\cdot\! \Big(1 \! - \! \lambda  \! + \! 
	\frac{\lambda}{m} \sum_{j=1}^{m} p_{2,U,t+j}^{(i)}  \!\cdot\!  p_{2,L,t+j}^{(i)} \Big). 
	\end{align}   
	
	\section{Numerical Results}
	In this section, we first present the BP decoding threshold ${\varepsilon_{_{\text{BP}}}}$ for some PIC-dTC ensembles by using the DE analysis derived in Section \ref{sec:de}. After that, we show the simulation results on the error performance of our PIC-dTCs. 
	We consider TC1 with ${\mathbf{G}_{\text{RSC}1}=\left [{\begin{matrix} 1 ~ \frac{5}{7} \end{matrix}}\right ]}$ as the benchmark, and we use TC2 with ${\mathbf{G}_{\text{RSC}2}= \left [{\begin{smallmatrix} 1 & 0 & \frac{5}{7} \\ 0 & 1 & \frac{3}{7} \end{smallmatrix}}\right ]}$ as component code to construct the PIC-dTCs ensembles. 
	
	\subsection{Density Evolution Results} \label{sec:de_result}	
	\begin{figure}[t]
		\centering
		\includegraphics[width=0.5\textwidth]{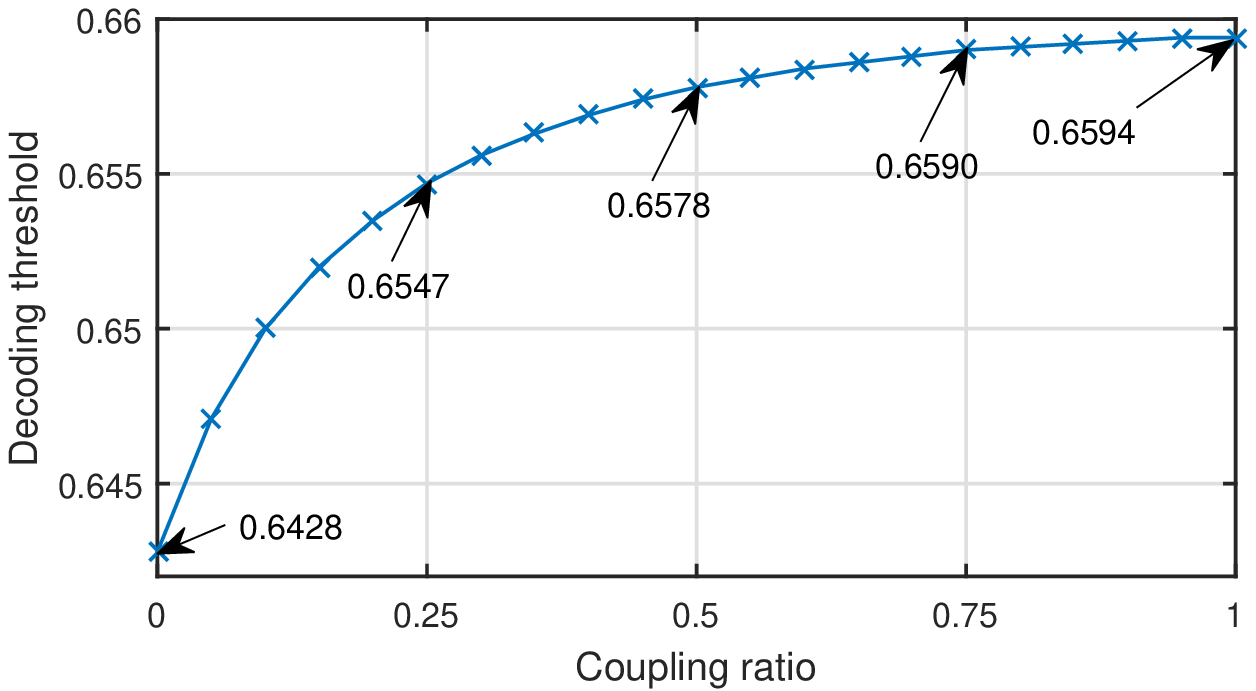}	
		\caption{BP Decoding thresholds of rate-$\frac{1}{3}$ PIC-dTCs with ${m=1}$ over coupling ratio $\lambda$.} 
		\label{fig:dec_th}	
	\vspace{2mm}
	\includegraphics[width=0.5\textwidth]{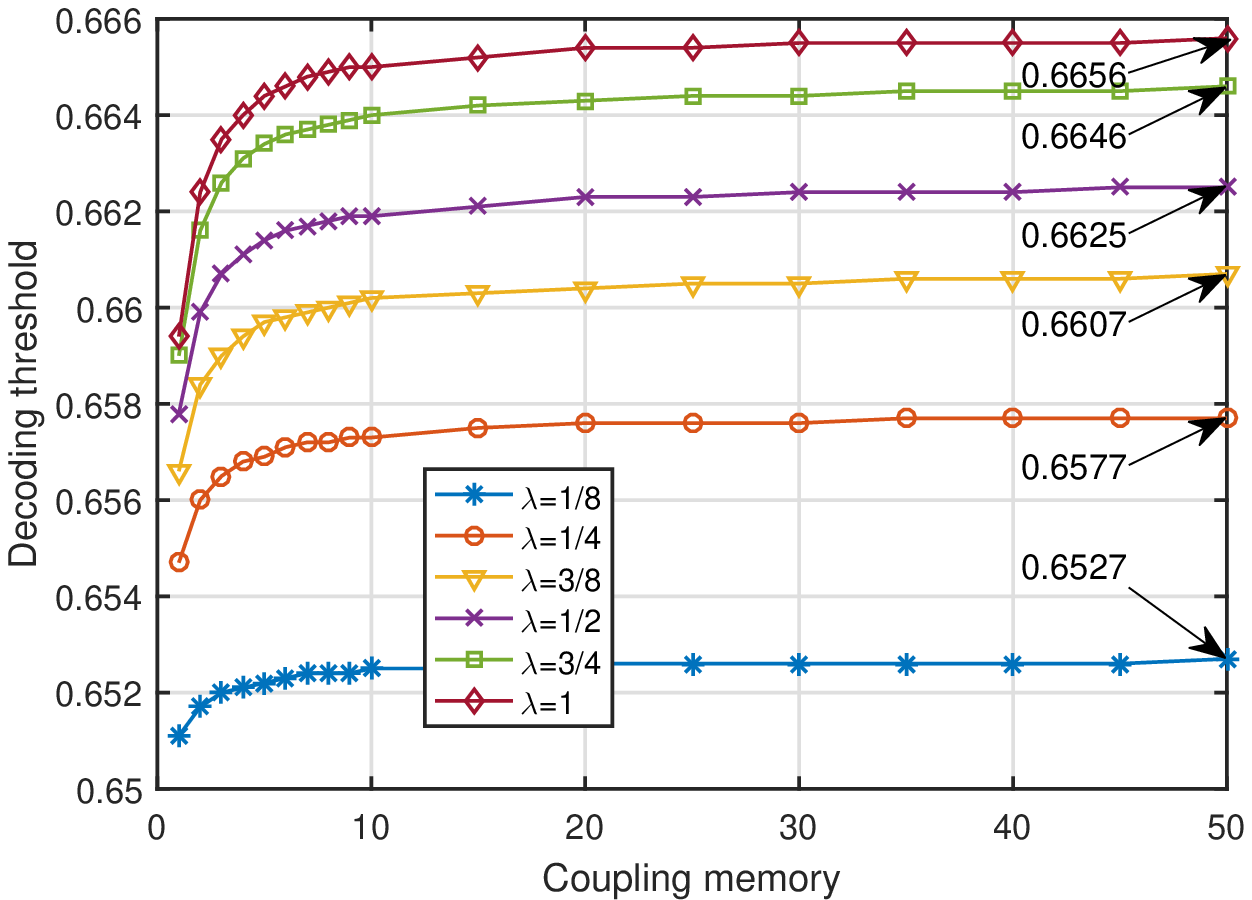}	
	\caption{BP decoding thresholds of rate-$\frac{1}{3}$  PIC-dTCs over coupling memory $m$.}
	\label{fig:dec_th_m}	
	\vspace{-5mm}
\end{figure}
	
	We plot the BP decoding thresholds of the PIC-dTCs versus the coupling ratio $\lambda$ for coupling memory ${m=1}$ in Fig. \ref{fig:dec_th}. It can be seen that the decoding threshold improves with $\lambda$ increasing.
	When ${\lambda=\frac{1}{2}}$, the gap between the decoding threshold and the BEC capacity is around 0.009. When $\lambda$ approaches $1$, the gap is only 0.0073. 
	However, it can also be observed that the increment of decoding threshold slows down when $\lambda$ keeps increasing. When $\lambda$ is increased from $\frac{3}{4}$ to 1, the decoding threshold has minor improvement. 
	
	To see how the decoding threshold improves with the coupling memory $m$, we plot the decoding thresholds of the PIC-dTCs as functions of $m$ in Fig. \ref{fig:dec_th_m}. For all considered coupling ratio $\lambda$ shown in the legend, the decoding threshold significantly improves when we increase $m$ from from 1 to 50. It is observed that the PIC-dTC can approach the BEC capacity with a gap around 0.0011 with ${\lambda=1}$ and ${m=50}$.
	Although it is unclear how the decoding performance would be when $m$ approach infinite, the threshold still keeps improving even when ${m>50}$. For example, with ${\lambda=1}$, the decoding threshold of PIC-dTC ensemble is 0.6657 when ${m=200}$, which is better than that for ${m = 50}$.  The investigation for threshold saturation is left for our future work. 
	
	We also compare our PIC-dTCs to other benchmark coding schemes. Specifically, in Table \ref{tb:e_bp}, we show the decoding thresholds of the PIC-TCs with ${\lambda=\frac{1}{2}}$, PIC-dTCs with ${\lambda=1}$, and state-4 SC-TCs in \cite{Moloudi-scTurbo}, including spatially coupled parallel concatenated convolutional codes (SC-PCCs), spatially coupled serial concatenated convolutional codes (SC-SCCs), and braided convolutional codes (BCCs). 
	Code rate ${R=\frac{1}{3}}$ and coupling memory ${m=1,5,50}$ are considered. Note that for ${m=50}$, we list the MAP decoding thresholds of the SC-TCs  due to threshold saturation. 
	It can be seen that the decoding thresholds of the PIC-dTCs is better than that of the PIC-TCs. When ${m=1}$, the PIC-dTCs outperform SC-PCCs and SC-SCCs, but is slightly worse than the BCCs. When $m$ is large, the PIC-dTC with ${\lambda=1}$ has the largest decoding threshold among all the benchmark codes. 
	
	\begin{table}[]
		\centering
		\begin{tabular}{l c c c}
			 \hline \\[-6pt]
			Ensemble & $\varepsilon_{_{\text{BP}}}, m=1 $& $\varepsilon_{_{\text{BP}}}, m=5 $ & $\varepsilon_{_{\text{BP}}}, m=50 $\\[2pt] \hline \\[-6pt]
			PIC-TC, $\lambda=\frac{1}{2}$       & 0.6566      & 0.6625    & 0.6639 \\[2pt]
			PIC-dTC, $\lambda=1$       & 0.6594     & 0.6644    & 0.6656 \\[2pt]
			SC-PCC        & 0.6553     & 0.6553    & 0.6553 \\[2pt]  
			SC-SCC        & 0.6437     & 0.6654    & 0.6654 \\[2pt]  
			BCC Type-I  & 0.6609     & 0.6650    & 0.6653 \\[2pt] 
			BCC Type-II & 0.6651     & 0.6653    & 0.6653 \\[2pt] \hline
		\end{tabular}
		\caption{BP Decoding thresholds of rate-$\frac{1}{3}$ spatially coupled codes. }
		\label{tb:e_bp}
	\end{table}
	
	\subsection{Error Performance Simulation Results}
	We now present simulation results for the PIC-dTCs with ${m=1}$ and code rate $\frac{1}{3}$. 
	We set ${L=100}$ to minimize the rate loss due to coupling termination.
	The error performance of PIC-dTCs is measured in terms of bit erasure rate (BER) versus the erasure probability of the BEC. 
	
\begin{figure}[t]
		\vspace{-2mm}
		\centering
		\includegraphics[width=0.5\textwidth]{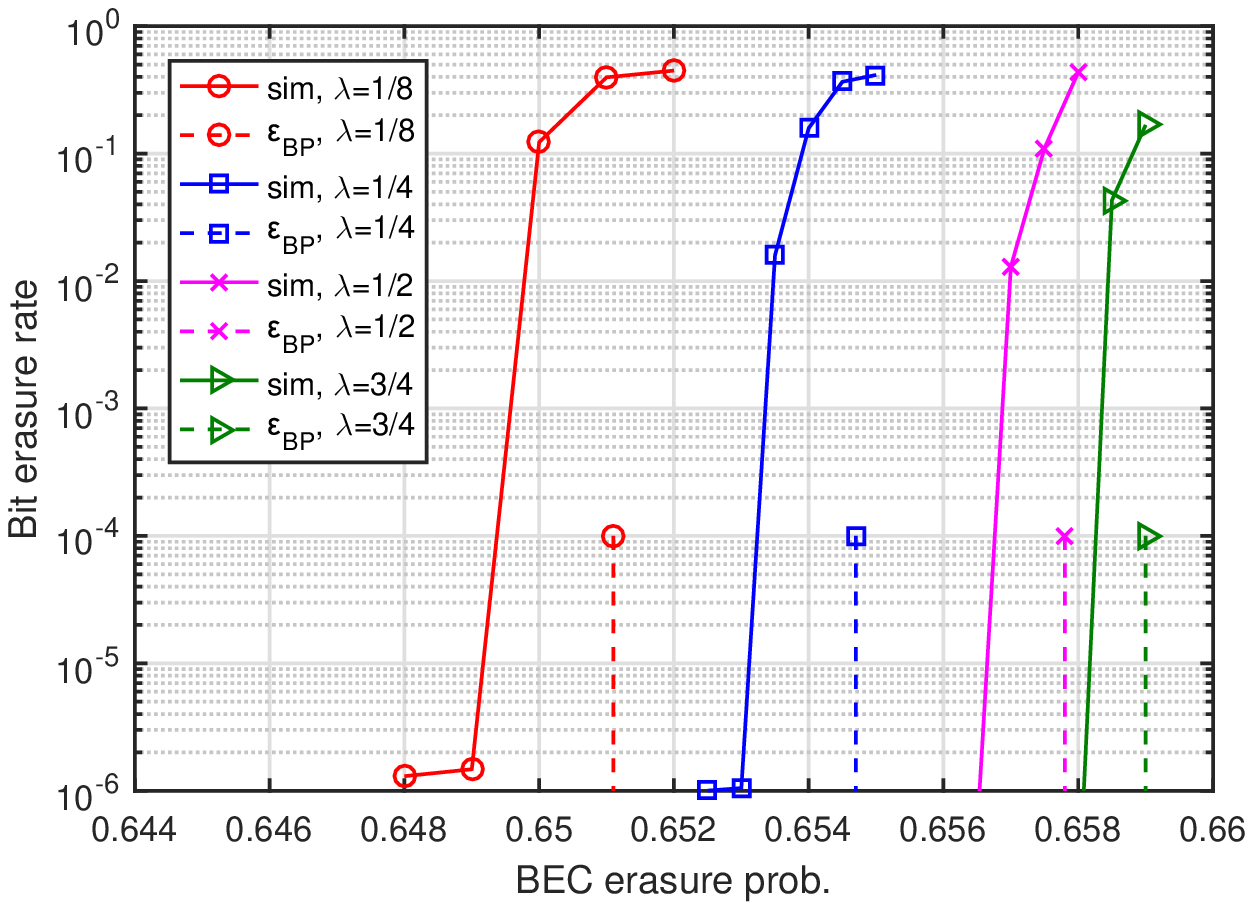}
		\caption{Error performance of rate-$\frac{1}{3}$ PIC-dTCs with ${m=1,}$ $K=10^5, L=100$.}
		\label{fig:ber_K100000}	
		\vspace{2mm}
		\includegraphics[width=0.5\textwidth]{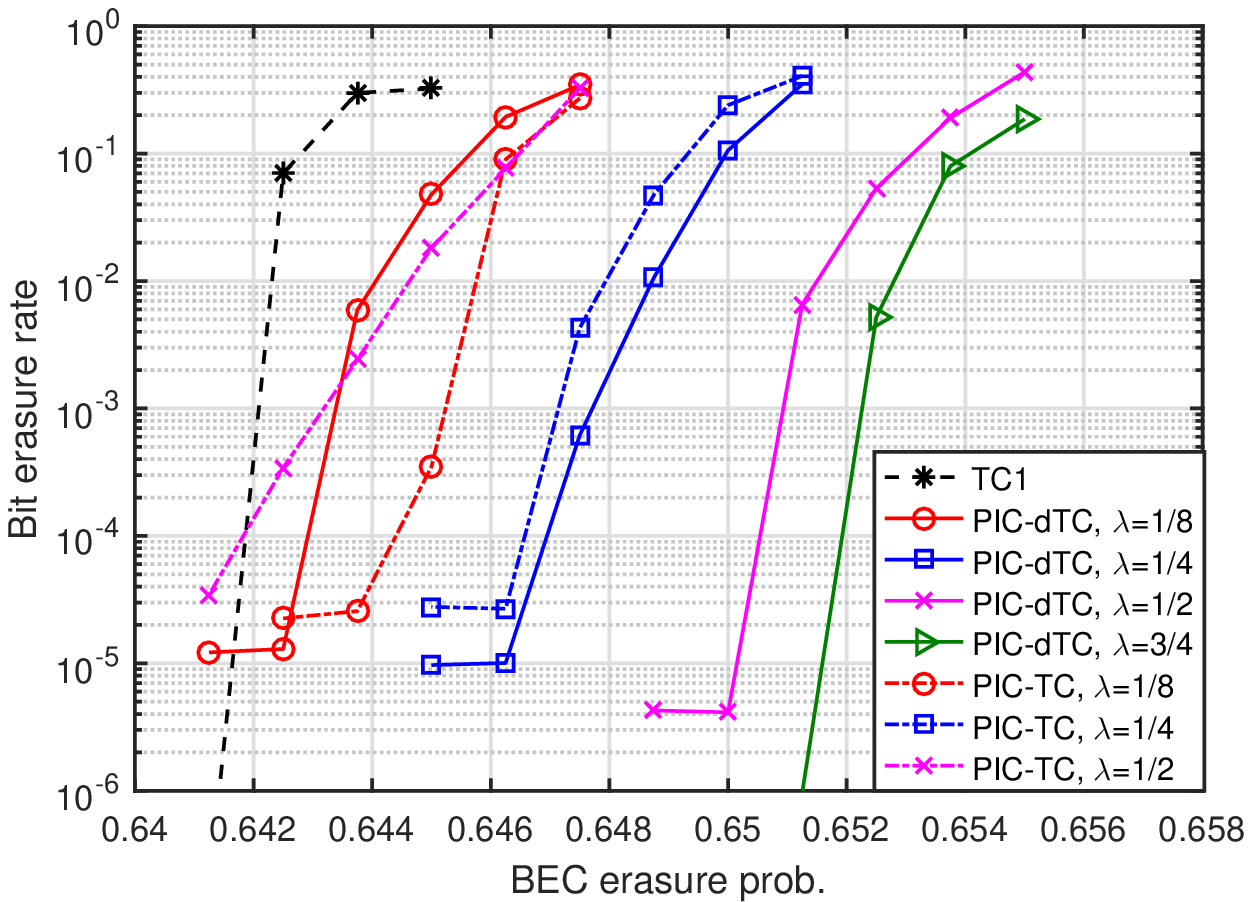}	
		\caption{Error performance of rate-$\frac{1}{3}$ PIC-dTCs and PIC-TCs with $m=1, K=10^4, L=100$.}
		\label{fig:ber_K10000}	
	\end{figure}

	In order to verify the correctness of the DE analysis, we compare the decoding threshold with the simulated BER (denoted as sim in the legend) of PIC-dTCs with ${K=10^5}$ and ${\lambda=\frac{1}{8},\frac{1}{4},\frac{1}{2},\frac{3}{4}}$. 
	The results are shown in Fig. \ref{fig:ber_K100000}. 
	It can be observed that all the PIC-dTCs are within 0.002 to the decoding threshold at a BER of $10^{-5}$. 
	
	In Fig. \ref{fig:ber_K10000}, to evaluate the performance of the PIC-dTCs with moderate information length, we plot the BER of PIC-dTCs with ${K=10^4}$, the PIC-TCs \cite{Yang-picTurbo,Qiu-picTurboDE} with ${K=10^4}$ (whose parity sequences are random punctured to increase the code rate to $\frac{1}{3}$), and TC1 with ${K=10^6}$. 
	It can be seen that all the PIC-dTCs outperform the uncoupled TC1 even though they have almost the same information length. 
	It is also observed that the PIC-dTCs can achieve better error performance by simply increasing $\lambda$, while the PIC-TCs need to optimize $\lambda$ in order to obtain a good error performance. Nevertheless, we observe that the PIC-dTCs have better error performance than the PIC-TCs when $\lambda$ is greater than $\frac{1}{4}$.
	
	\section{Conclusion}
	In this paper, we proposed the partially information coupled duo-binary turbo codes (PIC-dTCs), which do not have the rate loss as appeared in the PIC-TCs,  and investigated their performance. We considered the PIC-dTCs with coupling ratio ${\lambda \in [0,1]}$ and coupling memory ${m \geq 1}$. The encoding and decoding procedures for such codes are presented. 
	We then introduced the graph model representations of the PIC-dTCs ensembles, and derived the exact DE equations for any given coupling memory $m$ and coupling ratio $\lambda$. 
	We showed that the decoding thresholds of PIC-dTCs can approach the BEC capacity with a gap within 0.001 for a large coupling memory. The simulation results confirmed the correctness of the analysis. Both theoretical analysis and simulation results demonstrated the superior error performance of our PIC-dTCs over the uncoupled turbo codes and existing spatially coupled turbo codes at the same rate.
	
	 \section*{Acknowledgement}
	 The work was partially supported in part by the Australian Research Council Discovery Project under Grant DP190101363 and in part by the Linkage Project under Grant LP170101196.
	 
	\bibliographystyle{ieeetr}
	\bibliography{pic_turbo_isit_reference}

\begin{thebibliography}{10}

\bibitem{Felstrom-convLDPC}
A.~{Jimenez Felstrom} and K.~S. {Zigangirov}, ``Time-varying periodic
  convolutional codes with low-density parity-check matrix,'' vol.~45,
  pp.~2181--2191, Sep. 1999.

\bibitem{Kudekar-scLDPC}
S.~{Kudekar}, T.~J. {Richardson}, and R.~L. {Urbanke}, ``Threshold saturation
  via spatial coupling: Why convolutional {LDPC} ensembles perform so well over
  the bec,'' {\em {IEEE} Trans. Inf. Theory}, vol.~57, pp.~803--834, Feb 2011.

\bibitem{Lentmaier-scLDPC}
M.~{Lentmaier}, A.~{Sridharan}, D.~J. {Costello}, and K.~S. {Zigangirov},
  ``Iterative decoding threshold analysis for {LDPC} convolutional codes,''
  {\em {IEEE} Trans. Inf. Theory}, vol.~56, pp.~5274--5289, Oct 2010.

\bibitem{Xie-scLDPC}
Y.~{Xie}, L.~{Yang}, P.~{Kang}, and J.~{Yuan}, ``Euclidean geometry-based
  spatially coupled {LDPC} codes for storage,'' {\em {IEEE} J. Sel. Areas
  Commun.}, vol.~34, pp.~2498--2509, Sep. 2016.

\bibitem{Moloudi-scTurbo}
S.~{Moloudi}, M.~{Lentmaier}, and A.~{Graell i Amat}, ``Spatially coupled
  turbo-like codes,'' {\em {IEEE} Trans. Inf. Theory}, vol.~63, pp.~6199--6215,
  Oct 2017.

\bibitem{Moloudi-scHCC}
S.~{Moloudi}, M.~{Lentmaier}, and A.~{Graell i Amat}, ``Spatially coupled
  hybrid concatenated codes,'' in {\em SCC 2017; 11th International ITG
  Conference on Systems, Communications and Coding}, pp.~1--6, Feb 2017.

\bibitem{Huebner-LaminatedTC}
A.~{Huebner}, K.~S. {Zigangirov}, and D.~J. {Costello}, ``Laminated turbo
  codes: A new class of block-convolutional codes,'' {\em {IEEE} Trans. Inf.
  Theory}, vol.~54, pp.~3024--3034, July 2008.

\bibitem{Zhang-bcc}
W.~{Zhang}, M.~{Lentmaier}, K.~S. {Zigangirov}, and D.~J. {Costello}, ``Braided
  convolutional codes: A new class of turbo-like codes,'' {\em {IEEE} Trans.
  Inf. Theory}, vol.~56, pp.~316--331, Jan 2010.

\bibitem{Yang-picTurbo}
L.~Yang, Y.~Xie, X.~Wu, J.~Yuan, X.~Cheng, and L.~Wan, ``Partially
  information-coupled turbo codes for {LTE} systems,'' {\em {IEEE} Trans.
  Commun.}, pp.~1--1, 2018.

\bibitem{Yang-PIC_LDPC}
L.~Yang, Y.~Xie, J.~Yuan, X.~Cheng, and L.~Wan, ``Chained {LDPC} codes for
  future communication systems,'' {\em {IEEE} Commun. Lett.}, vol.~22,
  pp.~898--901, May 2018.

\bibitem{Wu-picPolar}
X.~{Wu}, L.~{Yang}, Y.~{Xie}, and J.~{Yuan}, ``Partially information coupled
  polar codes,'' {\em IEEE Access}, vol.~6, pp.~63689--63702, 2018.

\bibitem{Wu-picPolarBICM}
X.~{Wu} and J.~{Yuan}, ``Partially information coupled bit-interleaved polar
  coded modulation for 16-qam,'' in {\em 2019 IEEE Information Theory Workshop
  (ITW)}, pp.~1--5, 2019.

\bibitem{Qiu-picTurboDE}
M.~{Qiu}, X.~{Wu}, Y.~{Xie}, and J.~{Yuan}, ``Density evolution analysis of
  partially information coupled turbo codes on the erasure channel,'' in {\em
  2019 IEEE Information Theory Workshop (ITW)}, pp.~1--5, 2019.

\bibitem{Berrou-nonBinaryTurbo}
C.~{Berrou} and M.~{Jezequel}, ``Non-binary convolutional codes for turbo
  coding,'' {\em Electronics Letters}, vol.~35, pp.~39--40, Jan 1999.

\bibitem{Berrou-duoBinaryTurbo}
C.~{Douillard} and C.~{Berrou}, ``Turbo codes with rate-m/(m+1) constituent
  convolutional codes,'' {\em {IEEE} Trans. Commun.}, vol.~53, pp.~1630--1638,
  Oct 2005.

\bibitem{bcjr}
L.~{Bahl}, J.~{Cocke}, F.~{Jelinek}, and J.~{Raviv}, ``Optimal decoding of
  linear codes for minimizing symbol error rate (corresp.),'' {\em {IEEE}
  Trans. Inf. Theory}, vol.~20, pp.~284--287, March 1974.

\bibitem{turbo_yuan}
B.~{Vucetic} and J.~{Yuan}, {\em Turbo Codes Principles and Applications}.
\newblock Norwell, MA: Kluwer, 2000.

\bibitem{Kurkoski-turboDE}
B.~M. {Kurkoski}, P.~H. {Siegel}, and J.~K. {Wolf}, ``Exact probability of
  erasure and a decoding algorithm for convolutional codes on the binary
  erasure channel,'' in {\em IEEE Global Telecommunications Conference},
  vol.~3, pp.~1741--1745 vol.3, Dec 2003.

\end{thebibliography}
\end{document}